\documentclass[a4paper]{VisionStyle}
\usepackage{epsfig}

\def\simlt{\hbox{ \rlap{\raise 0.425ex\hbox{$<$}}\lower
0.65ex\hbox{$\sim$} }}
\def\ltorder{\hbox{ \rlap{\raise 0.425ex\hbox{$<$}}\lower
0.65ex\hbox{$\sim$}}}
\def\simgt{\hbox{ \rlap{\raise 0.425ex\hbox{$>$}}\lower
0.65ex\hbox{$\sim$} }}

\begin{document}

\title{{\sl Chandra} Observations of the Galactic Center and Nearby Edge-on Galaxies}

\author{Q. Daniel\,Wang} 

\institute{Department of Astronomy, University of Massachusetts, Amherst, MA 01003, USA}

\maketitle 

\begin{abstract}

I review our recent {\sl Chandra} surveys of the center region of the 
Milky Way and other nearby edge-on galaxies. Our Galactic center survey,
consisting of 30 overlapping pointings, provides an unprecedented 
high-resolution, broad-band X-ray view of a 0.8x2 
square degree swath along the Galactic plane. This X-ray panorama allows a 
separation of the discrete source and diffuse X-ray components.
Our preliminary analysis has led a detection of about 1000 discrete sources.
 Less than 20 of these sources are previously known objects, most of which are 
bright X-ray binaries. We find that the diffuse X-ray emission 
dominates over the contribution from faint
discrete sources and is globally associated with distinct
interstellar structures observed at radio and mid-infrared wavelengths.
We study how high-energy 
activities in the center region affect the immediate vicinity and may 
influence other aspects of the Galaxy. 

We have further observed nearby edge-on late-type disk galaxies in fields of 
low foreground Galactic extinction to gain external perspectives of the 
global disk/halo interaction. We have detected a giant diffuse 
X-ray-emitting corona around the galactic disk of \object{NGC~4631}. 
Extraplanar diffuse X-ray emission is also detected around \object{NGC~3556}.
These X-ray-emitting coronae morphologically resemble the 
radio halos of these galaxies, indicating a close connection between outflows 
of hot gas, cosmic rays, and magnetic field from the galactic disks.  
There is only marginal evidence for extraplanar diffuse X-ray emission 
in \object{NGC~4244} --- a galaxy with an extremely low star formation rate.
In general, the extraplanar diffuse X-ray emission is evidently
related to recent massive star forming activities in the 
galactic disks, especially in their central regions. 

\keywords{Missions: {\sl Chandra} -- galaxies: individual 
the Milky Way, \object{NGC~4631}, \object{NGC~4244}, \object{NGC~3556} (M108)}
\end{abstract}

\section{Introduction}

It is becoming increasingly clear that galaxies are dynamic ecosystems. 
Especially important in shaping the structure and evolution of galaxies
are high-energy activities that can generate high-velocity and high-temperature
gas. We have been carrying out a systematic observing program to gain various 
perspectives of high-energy phenomena and processes in galaxies. We are 
particularly interested in studying the production, transportation, and 
cooling of X-ray-emitting gas. 

In this presentation, I concentrate on describing our {\sl Chandra} surveys
of nearby edge-on galaxies and the center region of the Milky Way. 
At a distance of 8 kpc from the Sun, the Galactic center (GC) region 
represents the best laboratory for a detailed study of the population and 
nature of various discrete X-ray sources and the processes related to the gas 
heating and escaping in a galactic nuclear environment. Our nearby 
edge-on galaxy survey so far includes \object{NGC~4631}, \object{NGC~3556}, 
and \object{NGC~4244}, providing a global view of the galactic 
disk/halo interplay.
But before going on to discuss these new observations, let me 
briefly summarize what we know about X-ray-emitting gas around disk galaxies
in the pre-{\sl Chandra} era.

\section{Pre-{\sl Chandra} Observations}

{\sl ROSAT} observations gave only marginal evidence for the presence 
of extraplanar diffuse X-ray emission in several relatively normal edge-on 
spirals, most noticeably \object{NGC~4631} (\cite{qwang-E3:Wang95}) and NGC 891 
(\cite{qwang-E3:Breg97}). Both the spatial resolution and 
the counting statistics of these observations were very limited;
confusion with point-like sources was significant.

{\sl ROSAT} studies of the diffuse soft X-ray background have led to the 
conclusion that large amounts of X-ray-emitting gas are also present at high 
Galactic latitudes and beyond the Local Bubble around the Sun (e.g.,
\cite{qwang-E3:Snow97}). Although the Galaxy-wide distribution of the hot gas
is difficult to determine, it has been proposed 
that the GC region of the Milky Way may play an important role in 
supplying the hot gas (\cite{qwang-E3:Wang97}; \cite{qwang-E3:Almy00}; \cite{qwang-E3:Sofue00}). 

While the massive black hole at the dynamic center of our
Galaxy is not particularly active at present (e.g., \cite{qwang-E3:Bag01}), 
enhanced X-ray emission has
been observed in the surrounding region (e.g., \cite{qwang-E3:Koy96}; 
\cite{qwang-E3:Wang02}). The presence of various spectral lines (e.g.,
6.7-keV Fe XXV K$\alpha$),
as detected by {\sl ASCA}, has been taken as an indication for large 
amounts of very hot diffuse gas ($\sim 10^8$ K; \cite{qwang-E3:Koy96}). 
Such hot gas cannot be 
confined by the gravity of the region and is likely to escape. 

\begin{figure}[tbh]
\centerline{
}
\caption{{\sl ROSAT} all sky survey map of the inner region of the Galaxy
in the $\sim 1.5$ keV band (Snowden et al. 1997). The 
box in the middle marks the region covered by {\sl ROSAT} pointed
observations (Fig.~\ref{qwang-E3_fig2}).
}
\label{qwang-E3_fig1}
\end{figure}

\begin{figure}[tbh]
\centerline{
}
\caption{\protect\footnotesize
A close-up of the central region of the Galaxy.
The mosaic is constructed with {\sl ROSAT} PSPC observations 
(Sidoli et~al. 2001) 
in the highest energy band (1.5--2.4 keV; Snowden et al. 1997). 
The central rectangular box oriented along the
Galactic plane outlines the field mapped out by our {\sl Chandra} survey 
(Fig.~\ref{qwang-E3_fig3}). 
}
\label{qwang-E3_fig2}
\end{figure}

The outflow of the hot gas from the GC region may be 
responsible for much of the diffuse soft X-ray background, particularly 
in the inner field of the Galaxy. Fig.~\ref{qwang-E3_fig1} shows a large-scale
hollow-cone-shaped soft X-ray feature on each side 
of the Galactic plane. This distinct morphology
resembles those seen in nearby galaxies with active nuclear star
formation and suggests a hot gas outflow from the GC region of the
Milky Way. A close-up of the region, as presented in Fig.~\ref{qwang-E3_fig2},
further shows a bright soft X-ray plume that apparently connects to the
southern large-scale X-ray feature, $\simgt 300$ pc away from the Galactic plane
at the distance of the GC (Fig.~\ref{qwang-E3_fig1}). The exact morphology
of this plume is uncertain, because of the differential X-ray absorption 
across the field. A comparison of the {\sl ROSAT} PSPC and {\sl IRAS} 100 micron 
images suggests that the deficit of the soft X-ray emission between the 
plume and the circumnuclear region is most likely
due to X-ray shadowing by foreground interstellar dusty gas. 
There might also be outflows from other 
adjacent massive star forming regions. But the plume just below the 
GC is the most prominent and coherent vertical diffuse soft 
X-ray feature observed, and it may represent 
the hot gas outflow from the GC region to the southern halo of the Galaxy.

\section{{\sl Chandra} Survey of the Galactic Center}

\begin{figure*}[bht]
 \begin{center}
 \end{center}
\caption{
{\sl Chandra} ACIS-I intensity map of the Galactic central ridge. This 
map, constructed in the 1--8 keV range, is adaptively smoothed
with a signal-to-noise ratio of $\sim 3$ and is plotted logarithmically to 
emphasize low surface brightness emission. The saw-shaped boundaries of 
the map, plotted in Galactic coordinates, results from a specific roll 
angle of the observations.
}
\label{qwang-E3_fig3}
\end{figure*}

We have carried out a {\sl Chandra} survey of the Galactic ridge
around the GC (Fig.~\ref{qwang-E3_fig3}; \cite{qwang-E3:Wang02}). This survey consists
of 30 separate observations, which were all taken
in July 2001. A mosaic of these observations, including
only the data from the ACIS-I, covers a  
field of $\sim 0.8^\circ \times 2^\circ$. The survey data provide an 
invaluable database for studying the interplay 
between stars, gas, dust, gravity, and magnetic fields in the 
unique environment of the center region. 

\subsection{Discrete sources and Features}

Our preliminary analysis of the survey data, together with three deeper
archival observations, has led a detection of about 1000 discrete sources
(\cite{qwang-E3:Wang02}). 
Less than 20 of these sources are previously known objects, 
most of which are bright X-ray binaries. The newly-detected sources must 
represent a 
combination of various populations, including background AGNs and foreground 
stars as well as X-ray binaries (CVs, 
accreting neutron star and black hole systems) and massive stars in the
GC region, although their relative populations in the field
are still very much uncertain.

The number of the detected X-ray sources in our field
seems to be much higher than that expected from a measurement of the source
density in a relatively blank region of the Galactic plane (\cite{qwang-E3:Ebis01}). But, we are yet to quantify this source excess, 
taking into consideration such position-dependent parameters
as the detection threshold, the line-of-sight X-ray absorption, and the local
diffuse X-ray background level. The X-ray absorption, for example,
clearly varies across the field, affecting the surface brightness
distribution, particularly in the 1--3 keV energy band.  The majority
of the detected sources are in the energy range of 2--10 keV within
a luminosity range of $10^{32} - 10^{35} {\rm~ergs~s^{-1}}$ at the
distance of the GC. 

 \begin{figure}[!hb]
 \begin{center}
\psfig{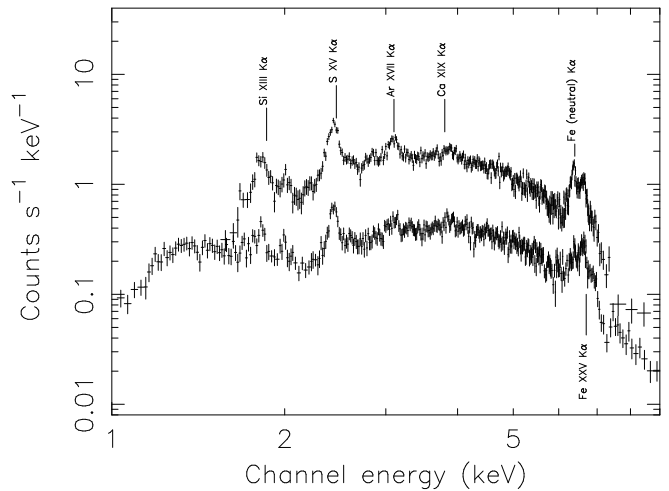}
 \end{center}
\caption{Comparison of accumulated 
source (lower) and diffuse X-ray emission (upper) spectra for the 
central enhancement above the surrounding background (see text). 
}
\label{qwang-E3_fig4}
\end{figure} 

Fig.~\ref{qwang-E3_fig4} compares the accumulated X-ray spectra of source and
diffuse components in the central region. These two spectra are extracted 
from an ellipse (with the major
and minor axis equal to 50$^\prime$ and $12^\prime$), 
centered on the Sgr A$^*$ and oriented along the Galactic
plane. Regions around the two brightest sources 
(1E 1740.7-2942 and 1E 1743.1-2843) are excluded to minimize the spectral 
pile-up problem. A diffuse background, obtained in the outer field and 
normalized in both exposure and area, is subtracted from the two spectra.
However, the X-ray surface brightness distribution varies strongly across the
field, especially in the region close to Sgr A.
Therefore, the above simple subtraction may still leave 
a considerable diffuse X-ray contribution (e.g.,
line emission) to the discrete source spectrum in Fig.~\ref{qwang-E3_fig4}. 

\begin{figure*}[!bht]
 \begin{center}
 \end{center}
\caption{Galactic center region across the spectrum: red: radio 90 cm (VLA;
LaRosa et al. 2001); green: mid-infrared (MSX; Price et al. 2001); blue: X-ray (1--8 keV; Fig.~\ref{qwang-E3_fig3})
}
\label{qwang-E3_fig5}
\end{figure*}

The total count rate from the detected discrete sources is comparable 
to that of the diffuse emission. But excluding the two brightest sources,
the source contribution is only $\sim 20\%$ of the diffuse emission 
from the central enhancement. Including the subtracted
X-ray background radiation, which is still uncertain, this fraction would be smaller.
No known source population with $L_x < 10^{32} 
{\rm~ergs~s^{-1}}$ could explain the diffuse hard X-ray emission as observed.

Our ongoing study of the X-ray spatial, spectral, and timing
properties as well as counterparts in other wavelength
bands will help us to constrain the nature of individual sources. 
Here I describe two 
examples of interesting X-ray sources and features that 
we have been studying.

\subsubsection{Arches cluster and its vicinity}

\begin{figure}[!bht]
 \begin{center}
 \end{center}
\caption{ACIS-I 1--7 keV intensity contours overlaid on the HST NICMOS
1.6 micro image of the Arches cluster (Figer et al. 1999). The contours
are at 26, 32, 44, 68, 120, 220, 420, 820, and 1620 $\times 10^{-3} {\rm~~counts~s^{-1}~arcmin^{-2}}$.}
\label{qwang-E3_fig6}
\end{figure}

The Arches cluster is the most compact massive star cluster
known in our Galaxy. 
The cluster was covered by one of our survey 
observations in a position that is significantly closer to the telescope 
axis than the {\sl Chandra} Cycle 1 image presented by Yusef-Zadeh et~al. 
(2002).  We find that the X-ray emission from the cluster consists of at 
least three discrete sources (Fig.~\ref{qwang-E3_fig6}); the apparently
extended northern source, for example, may contain multiple point-like 
components and/or a substantial amount of enhanced diffuse emission
in the vicinity.
The spectra of these sources show the prominent K$\alpha$ emission 
lines of highly-ionized ions such as S XV, Ar XVII, Ca XIX and Fe XXV
and are characterized by a two-temperature thermal plasma model 
of kT $\sim 1$ and 6 keV. Each source has an unabsorbed X-ray luminosity of 
$\sim 10^{35} {\rm~ergs~s^{-1}}$ in the 0.2--10 keV band. Portegies Zwart et al. (2001)
have argued that such high-luminosity X-ray sources may still represent 
colliding wind binaries of extremely massive stars. However, the centroids
of the X-ray sources don't match the positions of bright near-infrared stars, 
which typically have strong stellar winds (e.g., \cite{qwang-E3:Cot96}). 
Alternatively, the X-ray sources may represent emission peaks of the 
so-called cluster wind formed from the thermalization of colliding winds 
from a collection of individual stars (\cite{qwang-E3:Raga01}).

But the diffuse X-ray emission from
the vicinity of the cluster clearly has a different origin.
The X-ray spectrum of the emission
shows a strong 6.4-keV line with an equivalent width of 
$\sim 1$ keV  (\cite{qwang-E3:Yusef02}). We find 
that this K$\alpha$ emission of neutral to moderately-ionized irons
is spatially correlated with a dense 
molecular cloud observed in the region. Therefore, 
the diffuse X-ray emission may primarily represent the scattered  
X-ray radiation from the cluster.

\subsubsection{Nonthermal Radio Filament}

\begin{figure}[!tbh]
 \begin{center}
 \end{center}
\caption{X-ray contours overlaid on a 4.5 GHz radio continuum 
image of NTF G359.54+0.18 (Yusef-Zadeh 1997). 
The brightest point-like X-ray source in the field 
matches well spatially with a radio source.
}
\label{qwang-E3_fig7}
\end{figure}

One of the most unique high-energy phenomena observed in the GC region is the 
presence of numerous nonthermal filaments (NTFs) 
of radio emission (Fig.~\ref{qwang-E3_fig5}). While the origin of these filaments remains largely unknown, finding an X-ray counterpart can 
provide important constraints on the particle energetics.
We show in Fig.~\ref{qwang-E3_fig7} a convincing association of an X-ray 
``thread'' with NTF G359.54+0.18. This X-ray thread
is about $1^\prime$ long and has a width that is not adequately 
resolved on scales of $\sim 1^{\prime\prime}$. 
Interestingly, the X-ray emission is 
associated with the brightest of the two nearly parallel radio filaments
and appears to have a flat spectrum, consistent with a nonthermal 
nature of the thread.

\subsection{Large-scale diffuse X-ray emission}

The high spatial resolution of the {\sl Chandra} observations enables us,
for the first time, to measure the diffuse X-ray component in the GC region 
with little contamination from discrete sources (Fig.~\ref{qwang-E3_fig8}). 
The presence of prominent line emission from 
ions such as S XV, Ar XVII, and Ca XIX
(Fig.~\ref{qwang-E3_fig4}) in the diffuse X-ray spectrum indicates the presence \begin{figure*}[!bht]
 \begin{center}
 \end{center}
\caption{Diffuse X-ray emission across the X-ray spectrum: Red --- 1--3 keV; Green --- 3--5 keV;
Blue --- 5--8 keV. Source regions are replaced with 
interpolated 
values from surrounding regions. The resultant artifacts 
are apparent in regions around the two brightest sources
1E 1740.7-2942 and 1E 1743.1-2843. The images are adaptively smoothed with a Gaussian function
to achieve a local count-to-noise ratio of $\sim 8$; much of the small-scale 
fluctuations is due to the Poisson noise. 
}
\label{qwang-E3_fig8}
\end{figure*}

of gas at temperatures of $\sim 10^7$ K. The intensity image in the 1--3 keV 
band (Fig.~\ref{qwang-E3_fig8}), for example, is dominated by the line emission
(Fig.~\ref{qwang-E3_fig4}), 
which traces the diffuse hot gas component. 
The emission is greatly enhanced in the recent star forming region
close to Sgr A and the Radio Arc, where heating sources are abundant. 
The fast stellar winds from the three 
known young massive clusters alone (Galactic center, Arches, and Quintuplet;
\cite{qwang-E3:Figer99}), for example, account for a mechanical energy deposit
of order $10^{39}{\rm~ergs~s^{-1}}$, comparable to that from the central
cluster of the 30 Doradus nebula in the Large Magellanic Cloud (\cite{qwang-E3:Wang99}). 
The detectable lifetime of such clusters against the background 
star density is only 
a few megayears because of the strong gravitational tidal field in
the GC region (\cite{qwang-E3:Zwart02}). Therefore, the region may contain
a lot of massive stars, which will end their life as supernovae, heating the
ISM.

\begin{figure}[!hbt]
 \begin{center}
 \end{center}
\caption{6.4-keV line intensity contours overlaid on the HCN $J = 1 \rightarrow 0$ emission (Jackson et al. 1996) from the central region of the Galaxy.
The X-ray image is adaptively smoothed with a signal-to-noise ratio
of $\sim 3$, after discrete sources with count rates greater than
$10^{-3} {\rm~counts~s^{-1}}$ have been excised. A continuum contribution, 
estimated in the 4--6 keV and 7--9 keV bands, is subtracted.
}
\label{qwang-E3_fig9}
\end{figure}

However, the overall spectrum of the diffuse X-ray emission (Fig.~\ref{qwang-E3_fig4}) 
appears substantially harder than expected
for the thermal component alone. In fact, Fig.~\ref{qwang-E3_fig8} shows that 
the observed X-rays in different energy bands often arise in different regions,
indicating different origins. The prominent 6.4-keV line emission, for example,
is probably not related directly to the thermal component. Part of the line
emission is likely due to the fluorescent 
radiation from discrete sources, as in the case of the Arches cluster.
Alternatively, the emission may be produced by 
the filling of K-shell vacancies produced by low-energy cosmic
rays (\cite{qwang-E3:Val00}). For both scenarios, one might expect a good 
correlation between the 6.4-keV emission and dense molecular gas
tracers. However, 
such a correlation appears only globally, not on scales smaller than a 
few arcminutes (e.g., Fig.~\ref{qwang-E3_fig9}). This may indicate
that the 6.4-keV emission is strongly influenced by local sources of 
hard X-ray radiation and/or low-energy cosmic rays.

Furthermore, we find that the emission distribution in the 4--6 keV band, 
where no prominent 
emission line is present, is substantially different from 
those of the above two emission lines, although
differences in X-ray absorption are yet to be carefully 
considered. Most likely, the 4--6 keV band emission represents a combination
of the thermal hot gas, scattered discrete source emission, and 
radiation from additional processes such as bremsstrahlung of nonthermal
cosmic-ray electrons. The lack of enhanced X-ray emission from some of
the most prominent nonthermal radio filaments and mid-infrared features 
indicates that the inverse Compton scattering of the cosmic 
microwave background or the interstellar infrared radiation 
is not a significant contributor to the observed diffuse X-ray
emission.

Various radio studies have shown a global vertical configuration of 
inter-cloud magnetic fields, which extend to at least a few 
hundreds of pc away from the Galactic plane (e.g., \cite{qwang-E3:Lar00} 
and references therein). This configuration
could result from vertical outflows of magnetized materials from 
the circumnuclear region (e.g., \cite{qwang-E3:Wang01}). Indeed,
the diffuse X-ray emission appears to extend beyond our surveyed field 
(Fig.~\ref{qwang-E3_fig8}). The outflow of the hot plasma and relativistic 
particles 
from the GC region may also be responsible for larger scale diffuse radio and
soft X-ray features seen in the inner region of the Galaxy
 (\cite{qwang-E3:Wang97}; \cite{qwang-E3:Almy00}; \cite{qwang-E3:Sofue00}).

\begin{figure}[tbh]
 \begin{center}
 \end{center}
\caption{\object{NGC~4631} as seen edge-on
from the {\sl Chandra} ACIS-S in the 0.3--1.5 keV band (shown in blue and 
purple color) and Ultraviolet Imaging
Telescope (orange), tracing massive stars in the galaxy.
}
\label{qwang-E3_fig10}
\end{figure}

\begin{figure}[!htb]
 \begin{center}
 \end{center}
\caption{Central region of \object{NGC~4631} in the {\sl Chandra} ACIS-S 0.3--0.9 keV 
band  (shown in blue and purple color), compared with a {\sl Hubble} WFPC-2  
H$\alpha$ image (orange). 
}
\label{qwang-E3_fig11}
\end{figure}

\section{Nearby Edge-on Galaxies}

The best way to study the global galactic disk/halo interaction is
to observe nearby edge-on galaxies. We have observed with the {\sl Chandra} 
ACIS-S three nearby edge-on Scd galaxies as listed in Table 1. 
These galaxies, which show only little sign of nuclear starbursts,
  may be considered similar to our Galaxy in terms of morphology.
  Since they lie in directions of the sky with exceptionally low
  Galactic foreground absorption, they represent ideal cases to
  study their soft X-ray emission properties.
 Furthermore, while \object{NGC~4631} is strongly interacting 
with a companion, NGC3556 is an isolated 
galaxy. The star forming rate in \object{NGC~4244} is extremely low. The 
{\sl Chandra} observations allow for comparison of the 
relative importance of galaxy-galaxy interaction and internal galactic 
activity in generating extraplanar hot gas.

\begin{figure}[bht]
 \begin{center}
 \end{center}
\caption{Optical image with overlaid radio continuum contours of \object{NGC~3556} 
(King \& Irwin 1997).
}
\label{qwang-E3_fig12}
\end{figure}

We find only marginal evidence for diffuse X-ray emission in either the
disk or the halo of \object{NGC~4244}.
Both \object{NGC~4631} and \object{NGC~3556} show substantial diffuse soft X-ray emission
beyond stellar disks (\cite{qwang-E3:Wang01}; Figs.~\ref{qwang-E3_fig10}--\ref{qwang-E3_fig11};
Figs.~\ref{qwang-E3_fig12}--\ref{qwang-E3_fig13}). 
These observations provide the first unambiguous evidence for 
galactic coronae around galaxies that are similar to our Milky Way. 
The coronae morphologically resemble the radio halos of the galaxies 
(e.g., Fig.~\ref{qwang-E3_fig13}; \cite{qwang-E3:Wang95}; \cite{qwang-E3:Wang01}). One scenario for this radio-X-ray connection
is that hot gas outflows are accompanied by magnetic field 
and cosmic rays and are confined ultimately by magnetic field at large 
galactic radius (\cite{qwang-E3:Wang95}). 

\begin{table}[htb]
\caption{Parameters of Galaxies$^a$}
  \label{fauthor-E1_tab:tab1}
  \begin{center}
    \leavevmode
    \footnotesize
\begin{tabular}{lccc} 
 \hline \\[-5pt]
Parameter &\object{NGC~4631} & \object{NGC~3556} 		 & \object{NGC~4244}\\[+5pt]
\hline\\[-5pt]  
Incl. angle (deg)   &85 &86 &90\\
Distance (Mpc)  &6.9 &11.6 &3.1\\
$L_{\rm FIR}/D_{25}^2$  &3.5$\times 10^{40}$ &3.2$\times 10^{40}$ &0.29$\times 10^{40}$\\
exposure (ks) & 60 & 60 & 50\\
     \hline \\
      \end{tabular}
  \end{center}
$^a$Parameters are obtained from \cite*{qwang-E3:Col00} 
and \cite*{qwang-E3:Hoop99}. The mean surface far-infrared 
luminosity $L_{\rm FIR}/D_{25}^2$
is in units of ${\rm~ergs~s^{-1}~kpc^{-2}}$.
\end{table}

\begin{figure}[!bht]
 \begin{center}
 \end{center}
\caption{\object{NGC~3556} across the {\sl Chandra} ACIS-S spectrum: Red --- 
0.3--1.5 keV; Green --- 1.5--3 keV; Blue --- 3--7 keV. The radio contours 
are the same as in the previous figure.
}
\label{qwang-E3_fig13}
\end{figure}

The amount and extent of the extraplanar hot gas
clearly depend on the star formation rate of a galaxy.
For example, enhanced diffuse X-ray emission is apparently 
enclosed by numerous H$\alpha$-emitting loops blistered out from the central 
disk of \object{NGC~4631}, as is evident in a comparison with our deep
Hubble Space Telescope imaging (Fig.~\ref{qwang-E3_fig11}; \cite{qwang-E3:Wang01}). 

Expanding large-scale  shell-like HI structures have also been found 
in both NGC4631 and \object{NGC~3556} (\cite{qwang-E3:Rand93}; \cite{qwang-E3:King97}). These shells are very energetic; each requires an energy input equivalent to 
$\simgt 10^4$ SNe. Their nature is still a mystery. The two shells found in
\object{NGC~4631} are apparently associated with massive star forming regions,
as traced by HII regions. Thus, the energy may be provided by numerous
SNe and stellar winds from massive stars. Indeed, we find that two X-ray emission peaks spatially coincide with the shells. The peak associated with the
eastern shell is extended and may represent hot gas within the shell. 
The X-ray emission from the western peak is, however, dominated by 
a bright point-like source, presumably an X-ray binary. Thus, high spatial 
resolution X-ray observations are essential for determining the
nature of giant HI shells.

\begin{figure}[!tbh]
 \begin{center}
 \end{center}
\caption{{\sl Chandra} ACIS-S image of \object{NGC~4244} in the 0.3--1.5 keV band. The
contours outline the morphology of the optical disk of the galaxy. 
The straight lines outline the boundaries of the X-ray data coverage.
}
\label{qwang-E3_fig14}
\end{figure}

We find that the two partial HI shells  (the so-called eastern and western 
features) identified by \cite*{qwang-E3:King97} are not associated 
with any enhanced X-ray features. 
Therefore, these shells are probably not blown-out superbubbles 
produced by massive OB associations.

\section{Summary}

While our analysis of the {\sl Chandra} observations described above is 
still ongoing, I summarize some of the preliminary results as follows:

\begin{itemize}
 
\item We have obtained the first arcsecond-resolution X-ray panorama of the 
GC region, which allows for direct comparison with similar 
maps in radio and infrared. 

\item We have detected about 1,000 discrete X-ray sources in the GC region. 
A substantial fraction of these sources are likely to be
Galactic (e.g., massive stars and X-ray binaries).

\item We find that the diffuse X-ray emission dominates over the contribution 
from faint discrete sources. The spectrum of the diffuse X-ray emission 
indicates the presence of large amounts of hot gas with temperatures of
$\sim 10^7$ K. This hot gas originates in recent massive star forming regions 
and possibly in the central black hole itself. There is evidence that the 
hot gas is escaping the regions into the 
Galactic halo. The outflow of the hot gas may be responsible for large-scale
soft X-ray enhancements observed in the inner field of the Galaxy.

\item The fluorescence and reflection of the radiation from discrete sources 
may be important in explaining the overall hard spectrum of the 
diffuse X-ray emission. One example is the X-ray emission from the 
surrounding region of the Arches cluster. 
The filling of K-shell vacancies produced by non-relativistic cosmic-rays 
may also be important in producing Fe K-shell vacancies. 

\item We have detected a giant corona around the edge-on disk galaxy \object{NGC~4631} and a substantial diffuse extraplanar X-ray
component in \object{NGC~3556}. The diffuse X-ray morphology of these galaxies 
resembles their radio halos, indicating a close connection between outflows of hot gas, cosmic rays, and magnetic fields from the galactic disks. 

\item The extraplanar diffuse X-ray-emitting gas evidently originates in
star forming regions. In particular, enhanced
X-ray emission is apparently enclosed by numerous H$\alpha$-emitting 
loops blistered out from the central disk of \object{NGC~4631}. 

\end{itemize}

\begin{acknowledgements}

  I thank my collaborators for their contributions to the work described above 
and acknowledge the support from NASA/CXC through grants, 
GO0-1150, GO0-1095, GO1-2084, and GO1-2150.

\end{acknowledgements}
\vfil
\end{document}